\RequirePackage{fix-cm}

\documentclass[smallextended]{svjour3}       

\smartqed  

\usepackage[german,english]{babel}
\usepackage{graphicx}
\usepackage{siunitx}
\usepackage{amssymb}
\usepackage{epstopdf}
\usepackage{array}
\usepackage{soul}
\usepackage{mathptmx}      
\setuldepth{a}
\newcommand{\Ho}{$^{163}\textrm{Ho}$ }
\DeclareSIUnit{\sqrthz}{\ensuremath{\sqrt{\text{\hertz}}}}

\journalname{}

\begin{document}

\title{Specific Heat of Holmium in Gold and Silver at Low Temperatures}

\author{
	Matthew Herbst \and
	Andreas Reifenberger \and
	Clemens Velte \and
	Holger Dorrer \and
	Christoph E. D\"ullmann \and
	Christian Enss \and
	Andreas Fleischmann \and
	Loredana Gastaldo \and
	Sebastian Kempf \and
	Tom Kieck \and
	Ulli K\"oster \and
	Federica Mantegazzini \and
	Klaus Wendt	
}

\institute{Matthew Herbst \at
           Kirchhoff Institute for Physics, Heidelberg University, Im Neuenheimer Feld 227, 69120 Heidelberg, Germany \\
           \email{matthew.herbst@kip.uni-heidelberg.de}
           \and
           Andreas Reifenberger \at
           Kirchhoff Institute for Physics, Heidelberg University, Im Neuenheimer Feld 227, 69120 Heidelberg, Germany \\
           \email{andreas.reifenberger@kip.uni-heidelberg.de}
           \and
           Clemens Velte \at
           Kirchhoff Institute for Physics, Heidelberg University, Im Neuenheimer Feld 227, 69120 Heidelberg, Germany \\
           \email{clemens.velte@kip.uni-heidelberg.de}
           \and
           Holger Dorrer \at
           Institute of Nuclear Chemistry, Johannes Gutenberg University, Fritz-Strassmann-Weg 2, 55128 Mainz, Germany \\
           \and
           Christoph E. D\"ullmann \at
           Institute of Nuclear Chemistry, Johannes Gutenberg University, Fritz-Strassmann-Weg 2, 55128 Mainz, Germany \\
           GSI Helmholtzzentrum für Schwerionenforschung GmbH, Planckstr. 1, 64291 Darmstadt, Germany \\
           Helmholtz Institute Mainz, Johannes Gutenberg University, 55099 Mainz, Germany \\
           \and
           Christian Enss \at
           Kirchhoff Institute for Physics, Heidelberg University, Im Neuenheimer Feld 227, 69120 Heidelberg, Germany \\
           \and
           Andreas Fleischmann \at
           Kirchhoff Institute for Physics, Heidelberg University, Im Neuenheimer Feld 227, 69120 Heidelberg, Germany \\
           \and
           Loredana Gastaldo \at
           Kirchhoff Institute for Physics, Heidelberg University, Im Neuenheimer Feld 227, 69120 Heidelberg, Germany \\
           \and
           Sebastian Kempf \at
           Kirchhoff Institute for Physics, Heidelberg University, Im Neuenheimer Feld 227, 69120 Heidelberg, Germany \\
           \and
           Tom Kieck \at
           Institute of Nuclear Chemistry, Johannes Gutenberg University, Fritz-Strassmann-Weg 2, 55128 Mainz, Germany \\
           Institute of Physics, Johannes Gutenberg University, Staudingerweg 7, 55128 Mainz, Germany \\
           \and
           Ulli K\"oster \at
           Institut Laue-Langevin, 71 Avenue des Martyrs CS 20156, 38042 Grenoble, France \\
           \and
           Federica Mantegazzini \at
           Kirchhoff Institute for Physics, Heidelberg University, Im Neuenheimer Feld 227, 69120 Heidelberg, Germany \\
           \and
           Klaus Wendt \at
           Institute of Physics, Johannes Gutenberg University, Staudingerweg 7, 55128 Mainz, Germany \\
}

\date{ }

\maketitle

\begin{abstract}
The specific heat of dilute alloys of holmium in gold and in  silver plays a major role in the optimization of low temperature microcalorimeters with enclosed $^{163}\textrm{Ho}$, such as the ones developed for the neutrino mass experiment ECHo.
We investigate alloys with atomic concentrations of $x_\textrm{Ho}=\SI{0.01}{\percent} \text{ -- } \SI{4}{\percent}$ at temperatures between \SI{10}{\milli\kelvin} and \SI{800}{\milli\kelvin}. Due to the large total angular momentum $J=8$ and nuclear spin $I=7/2$ of $\textrm{Ho}^{3+}$ ions, the specific heat of \ul{Au}:Ho and \ul{Ag}:Ho depends on the detailed interplay of various interactions, including contributions from the localized 4f electrons and nuclear contributions via hyperfine splitting. This makes it difficult to accurately determine the specific heat of these materials numerically. Instead, we measure their specific heat by using three experimental set-ups optimized for different concentration and temperature ranges. The results from measurements on six holmium alloys demonstrate that the specific heat of these materials is dominated by a large Schottky anomaly with its maximum at $T\approx \SI{250}{\milli\kelvin}$, which we attribute to hyperfine splitting and crystal field interactions. RKKY and dipole-dipole interactions between the holmium atoms cause additional, concentration-dependent effects. 
With regard to ECHo, we conclude that for typical operating temperatures of  $T\leq\SI{20}{\milli\kelvin}$, silver holmium alloys with $x_\textrm{Ho}\gtrsim \SI{1}{\percent}$ are suited best.
\end{abstract}

\section{Introduction}
\label{chap:Introduction}

\Ho is an unstable holmium isotope which undergoes electron capture with the energy $Q_\textrm{EC} = (2.833 \pm 0.030_\textrm{stat} \pm 0.015_\textrm{syst})\,\textrm{keV}$ \cite{Eliseev2015} available for the decay. Based on its half-life of $\tau_{1/2}=4570\,\textrm{a}$, \SI{2e12}{} atoms of \Ho yield an activity of \SI{10}{\becquerel}\,\,\cite{Baisden1983}. For more than 30 years this nucleus is considered to be one of the best candidates to be used in experiments for the determination of the effective electron neutrino mass \cite{DeRujula1982,DeRujula2013,Laegsgaard1984,Hartmann1992,Gatti1997,Ranitzsch2012,Gastaldo2013}. The best approach for this kind of experiment is to perform a calorimetric measurement of the electron capture spectrum, as was proposed in 1983 by \textit{De R\'ujula} and \textit{Lusignoli} \cite{DeRujula1982,DeRujula2013}. Current technology is based on low temperature microcalorimeters \cite{Fleischmann2005} where \Ho atoms are enclosed in the particle absorber of the detector. Presently, two large experiments, namely ECHo \cite{Gastaldo2017} and HOLMES \cite{Alpert2015}, follow this approach.

As the name suggests, low temperature microcalorimeters are operated at temperatures below \SI{100}{\milli\kelvin}. They typically have dimensions in the order of a few hundred micrometers and a thickness of a few micrometers. In these detectors, the energy released by the decay of implanted \Ho leads to an increase in temperature, which is read out by a very sensitive thermometer. The temperature increase is proportional to the deposited energy and to the inverse of the total heat capacity of the detector. It follows that heat capacity is an important parameter for detector optimization.

In order to determine the heat capacity of the detector, it is necessary to gain a precise understanding of the contribution of the $\sim \! 10^{12}$ atoms of $^{163}$Ho, which are present in a dilute form in the absorber material. 
In commonly used hosts like gold and silver, holmium is present in the ionized state $\textrm{Ho}^{3+}$, which features a total electronic angular momentum of $J=8$ and a nuclear moment of $I=7/2$. Therefore, a non-negligible contribution due to magnetic interactions has to be expected. Surprisingly, this was not observed in a previous study on holmium-implanted gold films \cite{Prasai2013}, motivating us to perform this thorough analysis on the subject.

The work we present was performed as a part of the detector optimization for the ECHo experiment \cite{Gastaldo2017}. The ECHo experiment is designed to determine the effective electron neutrino mass using \Ho enclosed in the particle absorbers of metallic magnetic calorimeters (MMCs). As temperature sensor, MMCs utilize a paramagnetic material sitting in a constant magnetic field. The temperature-dependent Curie-like magnetization is then monitored using low-noise and high-bandwidth dc-SQUIDs \cite{Clarke2004}, which provide the voltage signal to be amplified and read out. In order to obtain the necessary \Ho source for the ECHo experiment, a prepurified and enriched $^{162}\textrm{Er}$ target is initially irradiated with thermal neutrons at the high flux reactor of the Institut Laue-Langevin (ILL). The thus created \Ho is then separated from the erbium via extraction chromatography \cite{Dorrer2018}. In order to precisely control the location of the holmium atoms, they are ionized and ion-implanted into the host. A mass separation step is added in between to assure high purity of the $^{163}$Ho and remove unwanted ionized species. This approach is used in ECHo and is also foreseen for HOLMES with some differences in the technological realization and, in particular, in the ion source \cite{Gallucci2019,Kieck2019}. 


Here we present two approaches to determine the specific heat of several dilute alloys. The first approach determines the holmium contribution to the heat capacity by comparing the signal size of two identical detectors, one with ion-implanted \Ho and one without. The second method is a calorimetric measurement of the specific heat based on the relaxation method \cite{Hwang1997}. As a host material, the ECHo collaboration plans to use a thin silver layer in future, instead of previously used gold. Thus, we performed measurements on both potential alloys.


\section{Theoretical Background}
\label{chap:TheoreticalBackground}

The heat capacity of a sample at constant volume is given by $C_V = (\partial U/\partial T)_V$, where $U$ symbolizes the internal energy. As the heat capacities at constant volume $C_V$ and at constant pressure $C_p$ are very similar for the samples considered here, we will not distinguish between $C_V$ and $C_p$ and omit the index. Since we are interested in the contribution of holmium to the heat capacity of gold and silver doped with holmium, we will in the following only discuss the intensive quantity $c_{\rm Ho} = (C-C_{\rm M})/N$, where $N$ denotes the total number of holmium ions in the respective sample and $C_{\rm M}$ (M = Au, Ag) the known electronic and phononic contributions of the undoped host material \cite{Martin1973}. In doing so, the major contribution due to internal degrees of freedom as well as minor changes of the sample's electronic or phononic contributions due to the holmium doping are mapped to the individual holmium atoms. Hence, even their interplay with the host material will be reflected adequately in $c_{\rm Ho}$.

Most relevant for the description of the specific heat of holmium in metallic hosts at temperatures below \SI{1}{\kelvin} are the hyperfine splitting and crystal field effects. As holmium concentrations in our samples are very low, we can describe the alloys as solid solutions and assume that the fcc structure of the host material is maintained \cite{Wunderlin1963}. The holmium occupies regular lattice sites and three electrons of the outer shells delocalize into the conduction band, resulting in Ho$^{3+}$ ions. According to Hund's rules, we therefore find $S =2$, $L =6$, and $J = 8$ for the spin, angular momentum, and total angular momentum quantum numbers. The partially filled inner 4f-shell is shielded by the fully occupied 5s- and 5p-shells and hence the resulting Hamiltonian accounting for hyperfine splitting is given by $H_\textrm{hf} \propto \,\vec I \cdot \vec J$, where $I = 7/2$ denotes the nuclear spin quantum number \cite{Lide,Abragam1970}. A hyperfine energy level splitting in the range of $\SI{0,3}{\kelvin} \cdot k_{\rm B}$ is expected \cite{Krusius1969}.

We should note here that a number of the results presented within this paper have been measured with $^{165}$Ho samples, which is naturally abundant allowing for the preparation of alloys by standard techniques, while the ECHo experiment uses $^{163}$Ho. As both isotopes have the identical nuclear spin $I$ and an almost identical nuclear moment ($^{163}$Ho: 4.23\,$\mu_{\rm N}$, $^{165}$Ho: 4.17\,$\mu_{\rm N}$) \cite{Lide}, we do not expect noticeable differences in the specific heat data discussed here.

Besides the hyperfine splitting, we need to consider crystal field effects. The charge distribution of the electronic 4f shell has a complex shape described by moments of higher order and hence interactions with electric field gradients of the lattice break the 17-fold degeneracy of the $J = 8$ ground state leading to a number of multiplets. Simulations indicate that the lowest excited crystal field multiplet is located below $\SI{1}{\kelvin} \cdot k_{\rm B}$ and hence is a second reason for a Schottky anomaly in our temperature region of interest \cite{Williams1969}. In fact, the hyperfine- and crystal field splitting cannot be separated from each other as they are coupled in $J$ and we expect a single, more complex Schottky peak. Specific heat measurements of pure holmium support this \cite{Krusius1969}.

Another observable effect originates from the pairwise interaction between hol\-mi\-um ions: Magnetic dipole-dipole interactions scale with $1/r^3$, where $r$ is the distance between the two holmium ions. Since the average distance $\hat{r}$ scales with $1/\sqrt[3]{x_{\rm Ho}}$, this interaction is concentration-dependent. As our data will show, contributions to the specific heat of a specific alloy will shift to lower temperatures for samples with lower concentration.
The same is true for the RKKY-interaction \cite{Ruderman1954,Kasuya1956,Yosida1957}, which also scales with $1/r^3$.
However, since this interaction is mediated indirectly by conduction electrons, its strength also depends on the host material. In the case of pure metals (Au, Ag) doped with erbium, the magnetic interactions have been shown to be relevant at temperatures below \SI{100}{\milli\kelvin} and are roughly $2 \textrm{--} 3$\, times stronger in silver than in gold \cite{Fleischmann2005}.

Based on this information, we investigate the influence of holmium concentration $x_\textrm{Ho}$, temperature, and host material on the specific heat of dilute holmium alloys. A brief summary of these parameters follows:
\paragraph{Holmium Concentration}\enskip 
In the ECHo experiment, each detector will contain about \SI{2e12}{} holmium ions corresponding to an activity of \SI{10}{\becquerel}. In the baseline design of the detector, the ions will be implanted in an area of $\SI{150}{\micro\meter} \times \SI{150}{\micro\meter}$, with simultaneous co-deposition of the host material in order to obtain a concentration of roughly \SI{1}{\percent}. The result of such an ion implantation is an inhomogeneous distribution of the holmium ions, with concentrations ranging from \SI{0}{\percent} to \SI{4}{\percent} \cite{Gamer2017}. For this reason, we investigated samples with atomic holmium concentrations between $x_{\rm Ho} = \SI{0.0162}{\percent}$ and \SI{4}{\percent}, in order to obtain information on the concentration dependence.  

\paragraph{Temperature}\enskip
In ECHo, the MMC detectors containing \Ho are operated at a temperature of about \SI{20}{\milli\kelvin}. We have performed heat capacity measurements between \SI{10}{\milli\kelvin} and \SI{800}{\milli\kelvin} in order to obtain a more detailed understanding of the different contributions to the specific heat. 
This broad temperature range allows us to precisely determine the position of the Schottky peak and characterize the tail towards low temperature.
\paragraph{Host Material}\enskip 
The initial choice for the host material for ECHo was gold, resulting in a number of studies of alloys of gold and holmium. However, the nuclear quadrupole moment of gold may affect the detector's performance. This has been observed for \ul{Au}:Er alloys \cite{Enss2000,Kempf2018}. As a result, new prototypes of MMCs for the ECHo experiment have recently been produced, in which \Ho was implanted into a thin silver layer (of the order of \SI{100}{\nano\meter}) grown on the original gold absorber layer. Here, we present a comparison of \ul{Au}:Ho and \ul{Ag}:Ho.


\section{Experimental}

\subsection{Samples}
\label{chap:samples}
We produced six samples, each based on gold (6N) or silver (5N) and doped with holmium at a (sub-)percent level. An overview appears in table \ref{tab:overview}. 

Sample 1 was made by implanting the electroplated gold absorber of a metallic magnetic calorimeter with $^{163}$Ho. The ion implantation was done at the \SI{30}{\kilo\volt} magnetic mass separator RISIKO facility in Mainz, Germany \cite{Schneider2016a}, however, without co-deposition of gold. The sample's holmium concentration was estimated by activity measurements (\SI{0.9}{\becquerel}) and through SRIM simulations \cite{SRIM} of the implantation profile \cite{Gamer2017}. As a result, we expect a spatially varying concentration ranging from \SI{0}{\percent} to \SI{4}{\percent} over an implantation depth of roughly \SI{10}{\nano\meter}.


All other samples S2 -- S6 were prepared by initial arc-melting in a clean (5N) argon plasma followed by multiple rotating and remelting steps, again in clean argon, to ensure a homogeneous holmium concentration throughout the sample. Then, samples were cut, pressed, sanded, and chemically etched to achieve the desired shape and mass. For these alloys we used the stable isotope $^{165}$Ho, which has a natural abundance of \SI{100}{\percent}. The holmium concentrations $x_{\rm Ho}$ of the gold-based samples S2 and S3 were determined by the mixing ratios of the source materials with an estimated accuracy of about \SI{10}{\percent}.
In order to obtain the holmium concentration of samples S4 -- S6, we performed magnetization measurements\footnote{MPMS XL-5 SQUID Magnetometer by Quantum Design, Inc., 10307 Pacific Center Court, San Diego, CA 92121, USA.} between \SI{2}{\kelvin} and \SI{300}{\kelvin}. With published crystal field parameters for holmium in silver (see section~\ref{chap:AgHo}), $x_{\rm Ho}$ could be determined within an estimated error of \SI{5}{\percent}.

\begin{table}
	\caption{Overview of the different samples S1 -- S6 investigated in this work. For each sample, the host material, its holmium concentration $x_{\rm Ho}$, and the platform used for the respective measurement are given.}
	\label{tab:overview}
	\begin{tabular}{cclc}
		\hline\noalign{\smallskip}
		Name&Host Material&$x_\textrm{Ho}\,[\%]$&Platform\\
		\noalign{\smallskip}\hline\noalign{\smallskip}
		S1 & Au & \num{0} -- \num{4} & 1 \\
		S2 & Au & \num{1.2}  & 2 \\
		S3 & Au & \num{0.12}  & 2 \\
		S4 & Ag & \num{1.66}  & 2 \\
		S5 & Ag & \num{0.184}  & 2 \\
		S6 & Ag & \num{0.0162} & 3 \\
		\noalign{\smallskip}\hline
	\end{tabular}	
\end{table}

\subsection{Gradiometric microcalorimeter}
A precise measurement of the contribution of implanted $^{163}$Ho to the total heat capacity of an MMC can be performed by monitoring the change of temperature of two identical pixels under the same deposition of energy, where one pixel is implanted with $^{163}$Ho and the other is not. The term pixel refers here to the thermal unit  consisting of an absorber (potentially with implanted holmium), a paramagnetic sensor layer, as well as smaller elements such as gold stems acting as thermal links. By comparing two such pixels, we benefit from the gradiometric layout of the underlying double meander, where both pixels are read out by the same SQUID \cite{Fleischmann2005,Fleischmann2009}. In the ECHo-1k chip, which was used here, seven such pixel pairs fulfill this requirement, in that only one of the two pixels has \Ho implanted in the gold absorber \cite{Gastaldo2017}. In the actual ECHo experiment the non-implanted pixel of these asymmetrically doped meanders are used for an in-situ background measurement during data acquisition.

\begin{figure}[t]
	\center
	\includegraphics[width=0.7\textwidth]{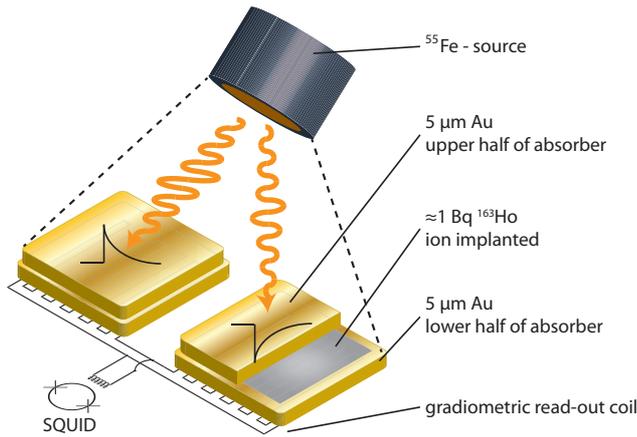}
	\caption{\label{fig:gradio} Schematic  drawing of platform 1, used to determine the specific heat of sample S1. Above the two gradiometrically wired, meander-shaped coils are the gold absorbers, one of which contains $^{163}$Ho. An $^{55}$Fe source provides K$_{\alpha}$ x-rays. Both the gold stems thermally linking the absorbers to the rest of the detector as well as the sensor layer between coils and absorbers are not shown for clarity.}
\end{figure}

One of the seven asymmetric pixel pairs present on the ECHo-1k detector chip was used for the heat capacity experiment. The \SI{5.89}{\kilo\electronvolt} $\textrm{K}_\alpha$ x-ray photons of an external $^{55}$Fe source were used to probe the thermal response of both pixels under investigation. Figure~\ref{fig:gradio} shows a schematic drawing of the experimental set-up.
The deposition of energy $\Delta E$ in one of the pixels leads to an increase in temperature $\Delta T\,=\,\Delta E/C$, where $C$ is the total heat capacity of this detector pixel. This change in temperature leads to a change in magnetization in the paramagnetic sensor material (\ul{Ag}:Er, not shown in figure~\ref{fig:gradio}). Due to flux conservation in the superconducting network containing the meander-shaped read-out coils\footnote{Magnetic fields caused by the persistent current in these coils are in the order of a few $\si{\micro\tesla}$ and do not have a measurable effect on the specific heat of the holmium \cite{Fleischmann2005,Velte}.} a compensating current is induced, which is read out using a current-sensing two-stage SQUID set-up operated in flux-locked-loop mode \cite{Clarke2004}. The resulting signal has the shape of a pulse, where the amplitude is proportional to the initial temperature increase of the detector and is thus also proportional to the deposited energy. As sketched in figure~\ref{fig:gradio}, the absorption of a photon in the left pixel will lead to a positive signal, while the absorption in the right pixel will lead to a signal of opposite polarity. The rise time depends on the electron-spin coupling in the sensor, while the decay time is defined by the ratio of the detector's heat capacity and the thermal conductance of the link to the heat bath. An additional contribution to the detector's heat capacity due to the implanted \Ho leads to a smaller signal size and to a larger decay time constant. 

\begin{figure}[t]
	\center
	\includegraphics[width=0.6\textwidth]{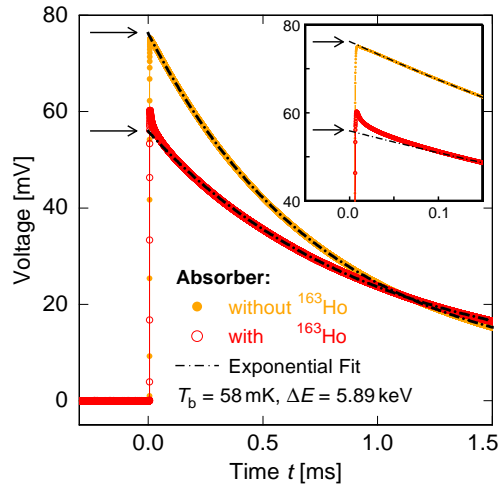}
	\caption{\label{fig:pulse} Comparison of the signal of a pixel without implanted holmium (yellow) and a pixel with implanted holmium (red) for an energy input of \SI{5.89}{\kilo\electronvolt} at T$_{\rm b} = \SI{58}{\milli\kelvin}$ originating from K$_{\alpha}$ photons of an external $^{55}$Fe source. Each curve is the result of averaging over roughly \num{1000} pulses. A clear difference in pulse height and decay time due to the implanted $^{163}$Ho is observable. In the inset, the two different pulse profiles at times below \SI{0.1}{\milli\second} become apparent. A comparison of the two signals allows us to determine the specific heat of sample S1.}
\end{figure}

Figure \ref{fig:pulse} shows the response to photon absorption in the two pixels of the discussed gradiometric detector at a temperature of \SI{58}{\milli\kelvin}. Each curve is the average of about \num{1000}~single K$_{\alpha}$ photon events. The signal corresponding to the detector without $^{163}$Ho appears in yellow, while the one corresponding to the detector with $^{163}$Ho appears in red. Since all the components of the two pixels are identical except for the $^{163}$Ho, the additional contribution due to the implanted \Ho ions to the total heat capacity of the detector causes the difference in the pulse profile. In particular, the Ho-implanted pixel shows a lower pulse amplitude and an additional exponential decrease for $t \lessapprox \SI{0.1}{\milli\second}$. A similar initial rapid decay has been previously reported in MMCs with \ul{Au}:Er \cite{Enss2000}, but is not yet fully understood. We observed in our case that it is more visible at higher bath temperatures, which suggests that it might be due to the larger Schottky contribution of $^{163}$Ho. Hence, we attribute this decay to internal relaxations within the holmium spin system. The portion of the red pulse after this initial steep decay contains the full information on the holmium subsystem. Thus, we extract a pulse height (arrows) fully sensitive to the holmium subsystem by extrapolating an exponential fit (black dot-dashed lines in figure~\ref{fig:pulse}) back to $t = \SI{0}{\milli\second}$. By experimental determination of the parameters of the read-out chain and the thermodynamic properties of the detector, a voltage to temperature conversion is calculated and the heat capacity of both pixels is extracted \cite{Velte,Ranitzsch2014}. In particular, we use the difference in pulse height to extract the heat capacity contribution of the \Ho ions and then normalize this by means of the measured activity (see section \ref{chap:samples}) and the known half-life of the decay.

\subsection{Direct Calorimetric Measurements}
In order to determine the specific heat of samples S2 -- S6 between \SI{20}{\milli\kelvin} and \SI{800}{\milli\kelvin}, we used two different set-ups based on the well-established relaxation meth\-od: The samples are placed on a platform using an adhesive ($\approx \SI{0.5}{\milli\gram}$ of Apiezon N grease\footnote{Apiezon Products, MI Materials Ltd, Hibernia Way, Trafford Park, Manchester M32 0ZD, United Kingdom}) and the time-resolved temperature response of the platform during the application of a well-defined heat pulse is monitored. A standard pulse-fitting method is applied to extract the heat capacity following the concept of \textit{Hwang} {\it et al}. \cite{Hwang1997}.

Platform~2, used for samples S2 -- S5, is based on a commercially available system\footnote{DR Heat Cap Puck QD-P107H by Quantum Design, Inc., 10307 Pacific Center Court, San Diego, CA 92121, USA.} with an addenda heat capacity of about \SI{4}{\nano\joule\per\kelvin} and a temperature resolution of \SI{1}{\micro\kelvin\per\sqrthz} at \SI{50}{\milli\kelvin}. Typically, we are limited to temperatures above \SI{30}{\milli\kelvin} with this platform. Further details of the set-up we recently summarized in \cite{Reifenberger2014}.

Platform~3 is based on a novel micro-fabricated chip with a \ul{Ag}:Er thermometer and SQUID readout that has been developed in house and is described in \cite{Reifenberger2020}. The platform features a very low addenda heat capacity of less than \SI{0.4}{\nano\joule\per\kelvin} and a temperature resolution of \SI{30}{\nano\kelvin\per\sqrthz} at \SI{50}{\milli\kelvin}, allowing us to measure samples with very low heat capacities. In particular, it enabled the measurement of sample S6.


\section{Results and Discussion}

\subsection{Specific Heat of \ul{Au}:Ho}
\label{ref:CResults}

\begin{figure}[t]
	\center
	\includegraphics[width=0.6\textwidth]{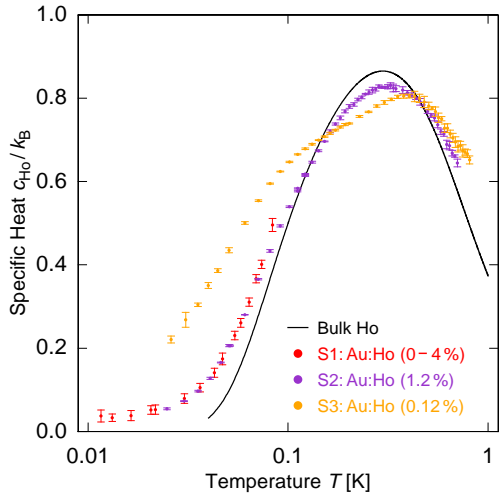}
	\caption{\label{fig:AuHo} Measured specific heat of samples S1, S2, and S3. The solid line represents the specific heat of bulk holmium.}
\end{figure}

The specific heat per holmium ion of the three \ul{Au}:Ho samples appears in figure \ref{fig:AuHo}, together with the published specific heat of bulk holmium \cite{Krusius1969,Lounasmaa1962}.
The indicated error bars for sample S1 are based on a comparison of eleven symmetric pixel pairs on the same ECHo-1k detector chip and their observed differences in pulse height for an identical energy input. Additionally, errors in the experimentally determined parameters used for voltage to temperature conversion were taken into account. The error bars of the other samples represent the statistical error of the average of typically 10 repeated measurements. Scaling errors for the individual data points due to the uncertainty in holmium concentration (as discussed in section~\ref{chap:samples}) are negligible for our analysis and not included.

The specific heat per holmium ion of bulk holmium is well understood \cite{Krusius1969}.
Its Schottky anomaly has a peak of roughly $0.9\,k_\textrm B$ at \SI{250}{\milli\kelvin}, which is almost entirely caused by magnetic hyperfine splitting in combination with crystal field effects, since the N\'eel temperature of holmium is \SI{133}{\kelvin} \cite{Koehler1966} and the 4f moments of the holmium ions cannot be thermally excited at the temperatures of interest, below \SI{1}{K}. 

We start the discussion of our results with sample S2 with a holmium concentration of \SI{1.2}{\percent}. The general shape of its specific heat curve resembles that of bulk holmium. At first glance, this is surprising given the difference in holmium concentration of two orders of magnitude. However, the similarity is understandable, since in both cases the concentration independent hyperfine splitting dominates the specific heat. Due to the close proximity of holmium ions, there are strong interactions between the 4f magnetic moments and the coupling strength suppresses almost all dynamics of the 4f magnetic moments.

A closer inspection reveals that the maximum of the Schottky anomaly is slightly reduced and occurs at somewhat higher temperatures with respect to bulk holmium. Since the effective crystal fields in the alloy \ul{Au}:Ho differ from bulk holmium, this can be understood qualitatively. The most notable difference, however, is that a large additional contribution to the specific heat appears at temperatures well below \SI{100}{\milli\kelvin}. We attribute this to specifically those holmium ions, whose coupling with neighbors are weak enough to contribute to the dynamics at these temperatures via RKKY and dipole-dipole interactions. Due to the random spatial distribution of ions, the interaction strengths vary, which leads qualitatively to the observed low temperature tail.

This explanation is supported by the fact that for sample S3, where the concentration is lowered by another order of magnitude, this contribution to the specific heat is further enhanced. Due to the reduced holmium concentration in this sample and the resulting increase of the average distance between holmium ions, a larger fraction of magnetic moments are no longer locked in interactions with their neighbors and can thus contribute to the dynamics. As also observed in sample S2, there is a slightly enhanced contribution compared to bulk holmium on the high temperature side of its maximum. In addition, the maximum itself is also shifted towards higher temperatures and lower values.

Of particular interest is the comparison with sample S1, since both the sample preparation and measurement technique are completely different. In addition, we are working with the radioactive holmium isotope $^{163}$Ho.
Despite these differences, we observe a remarkable quantitative agreement to the data of sample S2. This agreement has two important implications: First, it shows that possible defects generated by the implantation procedure play only a marginal role for the specific heat. Second, since platform~1 is sensitive on a timescale of \si{\micro\second}, while platform~2 determines the specific heat on a timescale of seconds to minutes, we conclude that the contributing degrees of freedom are the same and relax thermally within $\sim\!\SI{100}{\micro\second}$ (see red data in figure~\ref{fig:pulse}).

As mentioned in section~\ref{chap:samples}, the holmium concentration of sample S1 varies spatially between \SI{0}{\percent} and \SI{4}{\percent} \cite{Gamer2017}. The comparison with sample S2 indicates that with regards to the specific heat, the ion-implanted sample S1 behaves like an alloy with a concentration of $\sim\SI{1}{\percent}$. This suggests that a concentration of \SI{1}{\percent} is high enough to suppress 4f spin-flip contributions to the specific heat at $T<\SI{100}{\milli\kelvin}$, where the hyperfine and crystal field anomaly does not dominate.

\subsection{Specific Heat of \ul{Ag}:Ho}
\label{chap:AgHo}

\begin{figure}
	\center
	\includegraphics[width=0.6\textwidth]{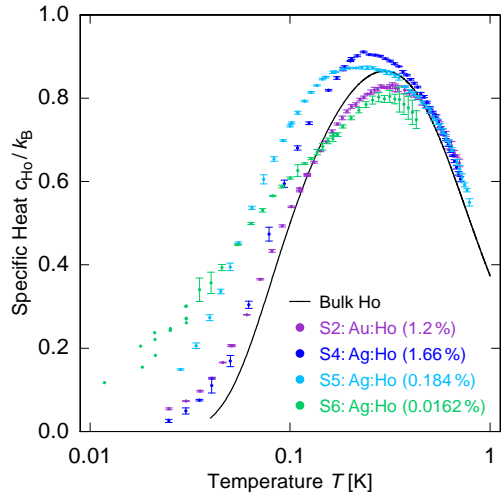}
	\caption{\label{fig:AgHo} Measured specific heat of samples S4, S5, and S6 with sample S2 and the bulk data for comparison. Note that for sample S6 there are insufficient data at low temperatures to show error bars, and individual data points appear instead.}
\end{figure}

Figure \ref{fig:AgHo} shows the specific heat per Ho ion for the \ul{Ag}:Ho samples S4, S5, and S6. 
For comparison, we also display the specific heat of bulk holmium and of sample S2. 

In silver holmium alloys we recognize many of the features we observed for holmium in gold: The high temperature region is almost independent of the concentration while at medium and at low temperatures we see qualitatively the same concentration-dependent effects. 

However, two differences between gold and silver as host material stand out. For that, we compare first samples S2 and S4, which have a similar concentration but a different host material. They show a difference in the location and height of the main Schottky peak. In order to understand this effect, we performed simulations for holmium ions in both gold and silver, based on a simplified model only including hyperfine splitting and crystal field effects\footnote{These simulations are based on unpublished work by S. H\"ahnle summarized in his master's thesis.}. Thus, the only way the two models differed from each other was in the crystal field parameters of their crystal field Hamiltonian. In particular, we used the crystal field parameters $W=-0.112$, $x=-0.357$ for \ul{Au}:Ho and $W=-0.373$, $x=-0.375$ for \ul{Ag}:Ho, which we obtained from converting the crystal field parameters from experimental results in literature \cite{Murani_1970,Williams1969,Stevens1952} into the $W$ and $x$ notation introduced in \cite{Lea1962}. Our simulations indicate that for \ul{Au}:Ho the maximum of the Schottky anomaly is shifted by \SI{25}{\milli\kelvin} towards higher temperatures and is $0.07\,k_\textrm B$ lower compared to \ul{Ag}:Ho. Since this matches almost perfectly with the observed differences between samples S2 and S4, we conclude that indeed the difference in the crystal field in the two host materials is responsible for these effects and may also be responsible for the shift with respect to bulk holmium.

The second observation is that already at lower holmium contents the con\-cen\-tra\-tion-dependent broadening of the Schottky peak in silver is equivalent to the broadening in gold, indicating a somewhat stronger RKKY interaction in silver. A similar trend was reported for Er ions in gold and silver \cite{Fleischmann2005}. 

\subsection{Implications for the ECHo Project}

Our analysis of the specific heat of dilute alloys of holmium in gold and silver results in a number of implications for the ECHo project. These relate to the three parameters mentioned in section \ref{chap:TheoreticalBackground}.

\paragraph{Holmium Concentration}\enskip 
The holmium concentration of the alloys has a direct impact on the specific heat of the material. At temperatures of $T\lesssim\SI{50}{\milli\kelvin}$, the specific heat per holmium ion of alloys with a high $x_\textrm{Ho}$ is lower than that of alloys with a lower $x_\textrm{Ho}$. Thus, a high concentration of holmium ions is preferable. Comparing sample S4 to bulk holmium indicates that a concentration $\gtrsim \SI{2}{\percent}$ does not yield further improvements regarding the low temperature specific heat. The results of the gold-based sample S2, on the other hand, still display a difference in height and slope, indicating that a further increase in $x_{\rm Ho}$ might be beneficial. This is a consequence of the stronger RKKY-interaction in silver-based alloys. These limits are reachable with current implantation techniques.

\paragraph{Temperature}\enskip 
We observe a large Schottky anomaly with a maximum at roughly $\SI{250}{\milli\kelvin}$ for all samples under investigation. The typical operating temperature of MMCs ($T\leq\SI{20}{\milli\kelvin}$) is in the regime where the contribution to the total heat capacity due to the presence of $^{163}$Ho ions is much smaller than at the peak. In addition, due to the good agreement between the measurement performed with the gradiometric microcalorimeter and the one performed through direct calorimetric measurements, we can state that contributions due to defects induced by the implantation process, if any exist, are marginal compared to the magnetic interactions of the holmium ions.
\paragraph{Host Material}\enskip 
Our experiments have demonstrated that the choice of host material affects both the peak of the Schottky anomaly at around \SI{250}{\milli\kelvin}, and the low temperature flank at $T\lesssim \SI{20}{\milli\kelvin}$. In general, silver seems to be preferable, since at low temperatures, the specific heat of silver alloys is lower than that of gold alloys with an identical holmium concentration. In order for the detector's absorbers to maintain good stopping power when using silver, a thin layer of \ul{Ag}:Ho may be surrounded with gold.

In addition to these three parameters regarding the specific heat of holmium alloys, our data allow us to draw conclusions regarding the total number of holmium ions which may be implanted into an ECHo absorber. This number is limited by the goal of keeping the heat capacity contribution of the implanted $^{163}$Ho ions below the heat capacities of the paramagnetic sensor and of the absorber material. Our measurements of the \ul{Au}:Ho sample S1 yield a heat capacity of \SI{3.3}{\pico\joule\per\kelvin} at \SI{20}{\milli\kelvin} for the non-implanted pixel of the gradiometric set-up. For the specific heat per $^{163}$Ho ion, we obtain a value of $0.05\,k_{\rm B}$ at $T=\SI{20}{\milli\kelvin}$ (see figure \ref{fig:AuHo}). The foreseen activity of the ECHo experiment is \SI{10}{\becquerel} per pixel, which is equivalent to \num{2e12} ions of $^{163}$Ho \cite{Gastaldo2017}. Thus, our measurements show that the implanted ions will increase the specific heat per pixel by only about \SI{42}{\percent}, which seems to be a good compromise between energy resolution and count rate in the present phase of the ECHo experiment.


\section{Conclusion}
The knowledge of the heat capacity of dilute alloys of holmium in host materials such as gold and silver is vital information for the optimization of the metallic magnetic calorimeters developed for the ECHo experiment. We performed heat capacity measurements of three \ul{Au}:Ho alloys with concentrations of \SI{1.2}{\percent}, \SI{0.12}{\percent}, and an implantation profile covering 0 to \SI{4}{\percent} and of three \ul{Ag}:Ho alloys with concentrations of \SI{1.66}{\percent}, \SI{0.184}{\percent}, and \SI{0.0162}{\percent}, respectively. For all samples we observed a large Schottky anomaly centered at about \SI{250}{\milli\kelvin} and reaching a height of about $0.9\,k_\textrm B$, similar to the one measured for bulk Ho. The shape of the Schottky anomaly depends on both the holmium concentration and the host material.

The results demonstrate that at the typical operating temperature of MMC detectors, the specific heat contribution of holmium ions is sufficiently low so that the ECHo experiment can be carried out with the foreseen activity of \SI{10}{\becquerel} per pixel. Additionally, our measurements demonstrate silver to be the more suitable host material in terms of heat capacity.


\section{Data Availability}
	The datasets generated during and/or analyzed during the current study are available in the Zenodo repository, doi.org/10.5281/zenodo.3585722.


\begin{acknowledgements}
Part of this research was performed in the framework of the DFG Research Unit FOR2202 “Neutrino Mass Determination by Electron Capture in 163Ho, ECHo” (funding under DU 1334/1-1 and DU 1334/1-2, EN 299/7-1 and EN 299/7-2, EN 299/8-1, GA 2219/2-1 and GA 2219/2-2). The research leading to these results has received funding from the European Union’s Horizon 2020 Research and Innovation Programme, under Grant Agreement no 824109. F. Mantegazzini acknowledges support by the Research Training Group HighRR (GRK 2058) funded through the Deutsche Forschungsgemeinschaft, DFG. We would like to thank George Seidel, Andreas Reiser, and Thomas Wolf for fruitful discussions and help in the preparation of the samples. Furthermore we thank the cleanroom team at the Kirchhoff Institute for Physics for technical support during device fabrication.
\end{acknowledgements}

\section*{Conflict of interest}
The authors declare that they have no conflict of interest.

\bibliographystyle{spphys}
\bibliography{library}

\begin{thebibliography}{10}
\providecommand{\url}[1]{{#1}}
\providecommand{\urlprefix}{URL }
\expandafter\ifx\csname urlstyle\endcsname\relax
  \providecommand{\doi}[1]{DOI \discretionary{}{}{}#1}\else
  \providecommand{\doi}{DOI \discretionary{}{}{}\begingroup
  \urlstyle{rm}\Url}\fi

\bibitem{Eliseev2015}
S.~Eliseev, K.~Blaum, M.~Block, S.~Chenmarev, H.~Dorrer, C.E. D{\"{u}}llmann,
  C.~Enss, P.E. Filianin, L.~Gastaldo, M.~Goncharov, U.~K{\"{o}}ster,
  F.~Lautenschl{\"{a}}ger, Y.N. Novikov, A.~Rischka, R.X. Sch{\"{u}}ssler,
  L.~Schweikhard, A.~T{\"{u}}rler, Phys. Rev. Lett. \textbf{115}, 062501 (2015)

\bibitem{Baisden1983}
P.A. Baisden, D.H. Sisson, S.~Niemeyer, B.~Hudson, C.~Bennett, R.~Naumann,
  Phys. Rev. C \textbf{28}, 337 (1983)

\bibitem{DeRujula1982}
A.~De~R\'ujula, M.~Lusignoli, Phys. Lett. B \textbf{118}, 429 (1982)

\bibitem{DeRujula2013}
A.~{De R\'ujula}.
\newblock {Two old ways to measure the electron-neutrino mass, arXiv:1305.4857}
  (2013)

\bibitem{Laegsgaard1984}
E.~Laegsgaard, J.U. Andersen, G.J. Beyer, A.~{De R\'ujula}, P.G. Hansen,
  B.~Jonson, H.L. Ravn, in \emph{Proceeding 7th Int. Conf. At. Masses Fundam.
  Constants} (1984)

\bibitem{Hartmann1992}
F.X. Hartmann, R.A. Naumann, Nucl. Instrum. Methods Phys. Res. A \textbf{313},
  237 (1992)

\bibitem{Gatti1997}
F.~Gatti, P.~Meunier, C.~Salvo, S.~Vitale, Phys. Lett. B \textbf{398}, 415
  (1997)

\bibitem{Ranitzsch2012}
P.C. Ranitzsch, J.~Porst, S.~Kempf, C.~Pies, S.~Sch{\"{a}}fer, D.~Hengstler,
  A.~Fleischmann, C.~Enss, L.~Gastaldo, J. Low Temp. Phys. \textbf{167}, 1004
  (2012)

\bibitem{Gastaldo2013}
L.~Gastaldo, P.C. Ranitzsch, F.~von Seggern, J.~Porst, S.~Sch{\"{a}}fer,
  C.~Pies, S.~Kempf, T.~Wolf, A.~Fleischmann, C.~Enss, A.~Herlert, K.~Johnston,
  Nucl. Instrum. Methods Phys. Res. A \textbf{711}, 150 (2013)

\bibitem{Fleischmann2005}
A.~Fleischmann, C.~Enss, G.M. Seidel, in \emph{Cryogenic Particle Detection},
  ed. by C.~Enss (Springer, Berlin/Heidelberg, 2005), p. 151

\bibitem{Gastaldo2017}
L.~Gastaldo, K.~Blaum, K.~Chrysalidis, T.~Day~Goodacre, A.~Domula, M.~Door,
  H.~Dorrer, C.E. D{\"u}llmann, K.~Eberhardt, S.~Eliseev, C.~Enss, A.~Faessler,
  P.~Filianin, A.~Fleischmann, D.~Fonnesu, L.~Gamer, R.~Haas, C.~Hassel,
  D.~Hengstler, J.~Jochum, K.~Johnston, U.~Kebschull, S.~Kempf, T.~Kieck,
  U.~K{\"o}ster, S.~Lahiri, M.~Maiti, F.~Mantegazzini, B.~Marsh, P.~Neroutsos,
  Y.N. Novikov, P.C. Ranitzsch, S.~Rothe, A.~Rischka, A.~Saenz, O.~Sander,
  F.~Schneider, S.~Scholl, R.X. Sch{\"u}ssler, C.~Schweiger, F.~Simkovic,
  T.~Stora, Z.~Sz{\"u}cs, A.~T{\"u}rler, M.~Veinhard, M.~Weber, M.~Wegner,
  K.~Wendt, K.~Zuber, Eur. Phys. J. Spec. Top. \textbf{226}, 1623 (2017)

\bibitem{Alpert2015}
B.~Alpert, M.~Balata, D.~Bennett, M.~Biasotti, C.~Boragno, C.~Brofferio,
  V.~Ceriale, D.~Corsini, P.K. Day, M.~{De Gerone}, R.~Dressler, M.~Faverzani,
  E.~Ferri, J.~Fowler, F.~Gatti, A.~Giachero, J.~Hays-Wehle, S.~Heinitz,
  G.~Hilton, U.~K{\"{o}}ster, M.~Lusignoli, M.~Maino, J.~Mates, S.~Nisi,
  R.~Nizzolo, A.~Nucciotti, G.~Pessina, G.~Pizzigoni, A.~Puiu, S.~Ragazzi,
  C.~Reintsema, M.~Ribeiro~Gomes, D.~Schmidt, D.~Schumann, M.~Sisti, D.~Swetz,
  F.~Terranova, J.~Ullom, Eur. Phys. J. C \textbf{75}, 112 (2015)

\bibitem{Prasai2013}
K.~Prasai, E.~Alves, D.~Bagliani, S.~Basak~Yanardag, M.~Biasotti, M.~Galeazzi,
  F.~Gatti, M.~Ribeiro~Gomes, J.~Rocha, Y.~Uprety, Rev. Sci. Instrum.
  \textbf{84}, 083905 (2013)

\bibitem{Clarke2004}
J.~Clarke, A.I. Braginski (eds.), \emph{{The SQUID Handbook: Fundamentals and
  Technology of SQUIDs and SQUID Systems}} (Wiley-VCH, Weinheim, 2004)

\bibitem{Dorrer2018}
H.~Dorrer, K.~Chrysalidis, T.D. Goodacre, C.E. D{\"{u}}llmann, K.~Eberhardt,
  C.~Enss, L.~Gastaldo, R.~Haas, J.~Harding, C.~Hassel, K.~Johnston, T.~Kieck,
  U.~K{\"{o}}ster, B.~Marsh, C.~Mokry, S.~Rothe, J.~Runke, F.~Schneider,
  T.~Stora, A.~T{\"{u}}rler, K.~Wendt, Radiochim. Acta \textbf{106}, 535 (2018)

\bibitem{Gallucci2019}
G.~Gallucci, M.~Biasotti, V.~Ceriale, M.~{De Gerone}, M.~Faverzani, E.~Ferri,
  F.~Gatti, A.~Giachero, P.~Manfrinetti, A.~Nucciotti, A.~Orlando, A.~Provino,
  A.~Puiu, J. Low Temp. Phys. \textbf{194}, 453 (2019)

\bibitem{Kieck2019}
T.~Kieck, H.~Dorrer, C.E. D{\"{u}}llmann, V.~Gadelshin, F.~Schneider, K.~Wendt,
  Nucl. Instrum. Methods Phys. Res. A \textbf{945}, 162602 (2019)

\bibitem{Hwang1997}
J.S. Hwang, K.J. Lin, C.~Tien, Rev. Sci. Instrum. \textbf{68}, 94 (1997)

\bibitem{Martin1973}
D.L. Martin, Phys. Rev. B \textbf{8}, 5357 (1973)

\bibitem{Wunderlin1963}
W.J. Wunderlin, B.J. Beaudry, A.H. Daane, {Institute of Metals Division - The
  Solid Solubility of Holmium in Copper, Silver and Gold}.
\newblock Tech. rep., The American Institute of Mining, Metallurgical, and
  Petroleum Engineers (1963)

\bibitem{Lide}
D.R. {Lide (Ed.)}, \emph{{CRC Handbook of Chemistry and Physics, Internet
  Version 2005}} (CRC Press, Boca Raton, 2005)

\bibitem{Abragam1970}
A.~Abragam, B.~Bleaney, \emph{{Electron Paramagnetic Resonance of Transition
  Ions}} (Clarendon Press, Oxford, 1970)

\bibitem{Krusius1969}
M.~Krusius, A.C. Anderson, B.~Holmstr{\"{o}}m, Phys. Rev. \textbf{177}, 910
  (1969)

\bibitem{Williams1969}
G.~Williams, L.L. Hirst, Phys. Rev. \textbf{185}, 407 (1969)

\bibitem{Ruderman1954}
M.A. Ruderman, C.~Kittel, Phys. Rev. \textbf{96}, 99 (1954)

\bibitem{Kasuya1956}
T.~Kasuya, Progr. Theor. Phys. \textbf{16}, 45 (1956)

\bibitem{Yosida1957}
K.~Yosida, Phys. Rev. \textbf{106}, 893 (1957)

\bibitem{Gamer2017}
L.~Gamer, C.E. D{\"{u}}llmann, C.~Enss, A.~Fleischmann, L.~Gastaldo, C.~Hassel,
  S.~Kempf, T.~Kieck, K.~Wendt, Nucl. Instrum. Methods Phys. Res. \textbf{854},
  139 (2017)

\bibitem{Enss2000}
C.~Enss, A.~Fleischmann, K.~Horst, J.~Sch{\"o}nefeld, J.~Sollner, J.S. Adams,
  Y.H. Huang, Y.H. Kim, G.M. Seidel, J. Low Temp. Phys. \textbf{121}, 137
  (2000)

\bibitem{Kempf2018}
S.~Kempf, A.~Fleischmann, L.~Gastaldo, C.~Enss, J. Low Temp. Phys.
  \textbf{193}, 365 (2018)

\bibitem{Schneider2016a}
F.~Schneider, K.~Chrysalidis, H.~Dorrer, C.E. D{\"{u}}llmann, K.~Eberhardt,
  R.~Haas, T.~Kieck, C.~Mokry, P.~Naubereit, S.~Schmidt, K.~Wendt, Nucl.
  Instrum. Methods Phys. Res. B \textbf{376}, 388 (2016)

\bibitem{SRIM}
J.F. Ziegler, M.D. Ziegler, J.P. Biersack, Nucl. Instrum. Methods Phys. Res. B
  \textbf{268}, 1818 (2010)

\bibitem{Fleischmann2009}
A.~Fleischmann, L.~Gastaldo, S.~Kempf, A.~Kirsch, A.~Pabinger, C.~Pies,
  P.~Ranitzsch, S.~Sch{\"{a}}fer, F.~von Seggern, T.~Wolf, C.~Enss, AIP
  Conference Proceedings \textbf{1185}, 571 (2009)

\bibitem{Velte}
C.~Velte, Measurement of a high energy resolution and high statistics
  $163\textrm{Ho}$ electron capture spectrum for the $\textrm{ECHo}$
  experiment.
\newblock Ph.D. thesis, Heidelberg University (2020)

\bibitem{Ranitzsch2014}
P.~Ranitzsch, {Development and characterization of metallic magnetic
  calorimeters for the calorimetric measurement of the electron capture
  spectrum of 163Ho for the purpose of neutrino mass determination}.
\newblock Ph.D. thesis, Heidelberg University (2014)

\bibitem{Reifenberger2014}
A.~Reifenberger, M.~Hempel, P.~Vogt, S.~Aswartham, M.~Abdel-Hafiez,
  V.~Grinenko, S.~Wurmehl, S.~Drechsler, A.~Fleischmann, C.~Enss, R.~Klingeler,
  J. Low Temp. Phys. \textbf{175}, 755 (2014)

\bibitem{Reifenberger2020}
A.~Reifenberger, A.~Reiser, S.~Kempf, A.~Fleischmann, C.~Enss, Rev. Sci.
  Instrum. \textbf{91}, 035118 (2020)

\bibitem{Lounasmaa1962}
O.V. Lounasmaa, Phys. Rev. \textbf{128}, 1136 (1962)

\bibitem{Koehler1966}
W.C. Koehler, J.W. Cable, M.K. Wilkinson, E.O. Wollan, Phys. Rev. \textbf{151},
  414 (1966)

\bibitem{Murani_1970}
A.P. Murani, J. Phys. C \textbf{3}, 153 (1970)

\bibitem{Stevens1952}
K.W.H. Stevens, Proc. Phys. Soc. A \textbf{65}, 209 (1952)

\bibitem{Lea1962}
K.R. Lea, M.J.M. Leask, W.P. Wolf, J. Phys. Chem. Solids \textbf{23}, 1381
  (1962)

\end{thebibliography}

\end{document}